\documentclass{emulateapj}
\usepackage{psfig}

\newcommand{\gsim}{\mbox{\hspace{.2em}\raisebox{.5ex}{$>$}\hspace{-.8em}\raisebox{-.5ex}{$\sim$}\hspace{.2em}}}
\newcommand{\lsim}{\mbox{\hspace{.2em}\raisebox{.5ex}{$<$}\hspace{-.8em}\raisebox{-.5ex}{$\sim$}\hspace{.2em}}}
\newcommand{\ssst}{\scriptscriptstyle}
\newcommand{\E}[1]{\times 10^{#1}}
\newcommand{\etal}{et al.}
       
\newcommand{\RA}[3]{{#1}^{{\rm h}}{#2}^{{\rm m}}{#3}^{{\rm s}}}
\newcommand{\decl}[3]{{#1}^{\circ}{#2}'{#3}''}

      \newcommand{\ps}{\,{\rm s}^{-1}}
\newcommand{\yr}{\,{\rm yr}}    \newcommand{\Msun}{M_{\odot}}
\newcommand{\cm}{\,{\rm cm}}    \newcommand{\km}{\,{\rm km}}
\newcommand{\parsec}{\,{\rm pc}}\newcommand{\kpc}{\,{\rm kpc}}
\newcommand{\ergs}{\,{\rm ergs}}        \newcommand{\K}{\,{\rm K}}
    \newcommand{\keV}{\,{\rm keV}}

\newcommand{\nel}{n_{e}}        \newcommand{\NH}{N_{\ssst\rm H}}
\newcommand{\no}{n_{\ssst 0}}

\newcommand{\nH}{n_{\ssst\rm H}}        \newcommand{\mH}{m_{\ssst\rm H}}
 
       \newcommand{\Einstein}{{\sl Einstein}}
\newcommand{\ROSAT}{{\sl ROSAT}} \newcommand{\ASCA}{{\sl ASCA}}
\newcommand{\Chandra}{{\sl Chandra}}
\newcommand{\du}{d_{4.3}}	\newcommand{\Du}{D_{20}}

\newcommand{\HI}{\ion{H}{1}}

\def\snr{{Kes~27}}

\shorttitle{}

\begin{document}

\title{The Thermal Composite
 Supernova Remnant Kesteven~27 as Viewed by {\sl CHANDRA}:
Shock Reflection from a Cavity Wall}
\author{
 Yang Chen\altaffilmark{1},
 Frederick D.\ Seward\altaffilmark{2},
 Ming Sun\altaffilmark{3}, and
 Jiang-tao Li\altaffilmark{1},
}
\altaffiltext{1}{Department of Astronomy, Nanjing University, Nanjing~210093,
       P.R.China}
\altaffiltext{2}{Harvard-Smithsonian Center for Astrophysics, 60 Garden Street,
       Cambridge, MA~02138}
\altaffiltext{3}{Department of Physics and Astronomy,
	Michigan State University, East Lansing, MI~48824}

\begin{abstract}
We present a spatially resolved spectroscopic study of the thermal composite
supernova remnant \snr\ with \Chandra.
The X-ray spectrum of \snr\ is characterized by K~lines
from Mg, Si, S, Ar, and Ca.
The X-ray--emitting gas is found to be enriched in sulfur and calcium.
The broadband and tricolor images show two incomplete
shell-like features in the northeastern half and brightness fading with
increasing radius to the southwest.
There are over 30 unresolved sources within the remnant.  None shows
characteristics typical of a young neutron star.
The maximum diffuse X-ray intensity coincides with a radio-bright
region along the eastern border.
In general, gas in the inner region is at higher temperature, and the
emission is brighter, than that in the outer region.
The gas in the remnant appears to be near ionization equilibrium.
The overall morphology can be explained by the evolution of the
remnant in an ambient medium with a density enhancement from west to east.
We suggest that the remnant was born in a preexisting cavity and that
the bright inner emission is due to the reflection of the initial
shock from the dense cavity wall. 
This scenario may provide a new candidate mechanism to explain
the X-ray morphology of other thermal composite supernova remnants.

\end{abstract}


\section{Introduction} \label{sec:intro}
Massive stars evolve rapidly and can explode not far from the dense
 clouds that were their
birthplaces.  The resulting supernova remnants (SNRs) thus are expected
to be in the vicinity of, and to interact with, molecular and/or \HI\ clouds.
Such interactions may account for the characteristics of
 the thermal composite (or mixed-morphology) SNRs, which radiate
bright thermal X-ray emission interior to their radio shells
and have faint X-ray rims (Green \etal\ 1997; Rho \& Petre 1998;
 Yusef-Zadeh et al.\ 2003).
Several mechanisms have been proposed to interpret the X-ray
morphology of the thermal composites.
These include (1) a radiatively cooled rim and a hot
interior (e.g., Harrus et al. 1997; Rho \& Petre 1998),
(2) a hot interior with density increased by thermal
conduction (Cox et al.\ 1999; Shelton et al.\ 1999),
(3) an increase in internal gas density due to evaporation
of the engulfed cloudlets (White \& Long 1991), and
(4) shock interaction at the edge
of a cloud but seen brightened in the projected interior
(Petruk 2001; as summarized in, e.g., Chen et al.\ 2004).
Recently, Shelton et al. (2004) have explained the X-ray
properties of SNR~W44 by invoking thermal conduction and bright metal 
emission due to dust destruction and ejecta enrichment in the 
interior.  Also,
a study based on ionization states of hot interior plasma
suggests that the thermal composite (or mixed-morphology)
phenomenon may be an evolutionary state of SNRs due to thermal
conduction (Kawasaki et al.\ 2005). 
The sample comprised six thermal composites including
\snr, which is the subject of this paper.

In none of these mechanisms has the influence of the supernova
progenitor on the surrounding medium been considered. 
Massive stars may sculpt a cavity with their energetic stellar
winds and ionizing radiation before they explode as core-collapse supernovae.
The observational effect of the SNR's blast wave 
impacting the cavity wall should be of particular interest.
After almost free expansion in the cavity, the blast wave will ``reflect''
from the cavity wall and a strong reverse shock will heat the interior
material.
Thus, a particularly X-ray bright interior may be expected in this scenario,
as we will outline for the case of \snr.

Kesteven~27 (G327.4$+$0.4) is one of the archetypical thermal composite
SNRs (Rho \& Petre 1998; Enoguchi et al.\ 2002).
As seen in X-rays with \Einstein\ (Seward 1990), \ROSAT\ (Seward et al.\ 1996),
and \ASCA\ (Enoguchi et al.\ 2002), it appears prominently brightened
in the center, especially along an inner, broken ring.
This is different from most other thermal composites, which
are centrally-peaked in X-rays.
In the radio band (Kesteven \& Caswell 1987; Whiteoak \& Green 1996)
there is a clear-cut rim, except in the southwest.
The radio emission is brightest in the east, with the surface brightness
peaking near the southeastern border, where the
blast wave may be striking a dense cloud. The radio emission gradually
 fades toward the southwest, where the remnant seems to break out
to a lower density region.
It was noted that the radio image displays multiple arcs centered
on the eastern half, from which a series of arclike shells may
have emanated (Milne et al.\ 1989).

Although OH masers, which usually mark shock interaction with molecular
clouds, are not detected in \snr\ (unlike
 other thermal composites such as 3C~391, W28, and W44)
(Green et al.\ 1997), this SNR is found to be embedded in an \HI\
cloud complex at a local standard of rest velocity of $70\km\ps$.
There is an \HI\ ridge or shell just exterior to the rim,
delineated by the radio continuum contours, except on the
west and southwest sides (McClure-Griffiths et al.\ 2001).
McClure-Griffiths et al.\
suggest that this confirms that the remnant shock
is impacting a density enhancement from the inner void.
The \HI\ observation provides a dynamical distance estimate of
$\sim4.3\kpc$, on the far side
of the Scutum-Crux arm.
In this paper, we take this value as a reference for the distance:
$d=4.3\du\kpc$.

We present a spatially resolved X-ray spectroscopic
study of the remnant using a \Chandra\ ACIS-I observation.
In \S~2, we briefly describe the observation and data calibration
and present our analysis and results.
The physical properties of the thermal X-ray emission
are discussed in \S~3.
We summarize our results in \S~4.
Statistical errors are all presented at the 90\% confidence level.

\section {Observations and Data Analysis}

\begin{figure}[tbh!] 
\centerline{ {\hfil\hfil
\psfig{figure=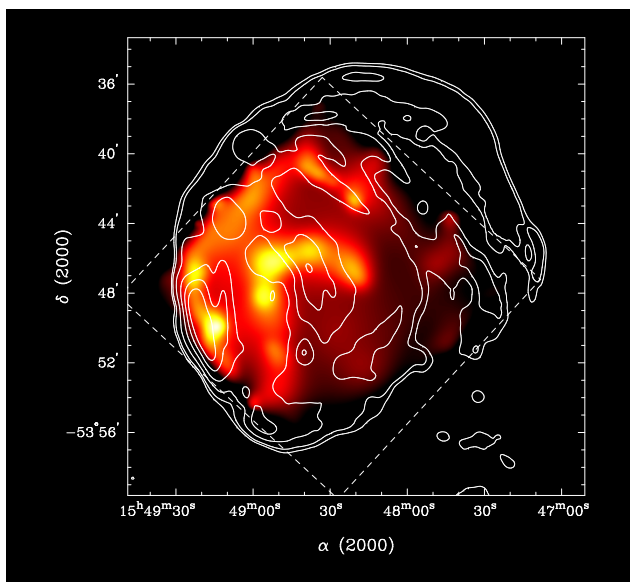,height=3.in,angle=270, clip=}
\hfil\hfil}}
\caption{
Broadband (0.3--7 keV) diffuse X-ray image (on a square root scale between
$64$ and $480$ photons$\cm^{-2}\ps\,\mbox{sr}^{-1}$)
overlaid with MOST 843~MHz radio
contours (at square root scale levels
0.01, 0.02, 0.05, 0.10, 0.17, 0.25, and 0.36
${\rm~Jy~beam^{-1}}$; from Whiteoak \& Green 1996).
The X-ray image is exposure-corrected and was
adaptively smoothed to achieve a S/N of 4
(using the CIAO program {\em csmooth}).
The dashed box denotes the field of view of ACIS-I.
}
\label{f:bband}
\end{figure}

\snr\ was observed with the Advanced CCD
Imaging Spectrometer (ACIS) on board the \Chandra\ X-Ray Observatory
on 2003 June 21 (ObsID 3852) for 39 ks.
The target center ($\RA{15}{48}{31.5}$, $\decl{-53}{46}{20}$; J2000) was placed
 for optimal coverage of the X-ray--bright region
of the remnant (see Fig.~\ref{f:bband}) on the four ACIS-I CCD chips.
We reprocessed the event files (from level 1 to level 2) 
 using the \Chandra\ Interactive Analysis of Observations ({\small CIAO})
 data processing software (ver.\ 3.2.2)
\footnote{http://cxc.harvard.edu/ciao}
to remove pixel randomization and to correct for CCD charge transfer
inefficiencies.
After removing flares with count rates greater than
1.2 times the mean light-curve value, a net exposure of 37 ks remained
and was used for analysis. 

\subsection{Source Identification and Optical Counterparts}

A major goal of this observation was to search for an internal
compact object.  There is obviously no bright Crab-like pulsar or 
pulsar wind nebula (PWN) within Kes 27.  However, less
energetic PSR/PWN examples have been observed in many
other remnants, for example, IC~443 (1.5 kpc
distant) and the Vela SNR (0.25 kpc distant) (Olbert et al. 2001;
Helfand et al. 2001).  Upper limits can be set for these types of PWNe.

If moved to the
more distant Kes 27, the IC 443 PWN ($L_x = 3 \times 10^{33}$ ergs
s$^{-1}$, $2^{\prime}\times 1^{\prime}$ in extent)
would appear as a $0.7^{\prime} \times 0.35^{\prime}$ patch in the 2--8 keV
X-ray band, with strength $\approx 20$ counts ks$^{-1}$.  We can set an
upper limit of 1 count ks$^{-1}$ for an object of this size in
the interior of Kes 27, so
$L_x \leq 1.5 \times 10^{32}$ ergs s$^{-1}$.  Similar reasoning for the
Vela PWN ($1.6^{\prime}$ in diameter, $L_x = 7 \times 10^{32}\ergs\ps$) 
predicts 7 counts ks$^{-1}$ from a
$6^{\prime\prime}$ region.  In this case we can set an
upper limit of 0.5 counts ks$^{-1}$, corresponding to 
$L_x \leq 0.5 \times 10^{32}$ ergs s$^{-1}$.  Note that an upper limit
depends strongly on the size of the object.
Leahy (2004) has
suggested that the PSR/PWN within the IC 443 boundary is really
associated with the nearer remnant G189.6+3.3.  Although this would
reduce its luminosity to about one-quarter of that quoted above, our upper
limit does not change.

The thermal composite remnant Kes 79 (7 kpc distant) contains an
unresolved central compact object (CCO) with $L_x = 7 \times 10^{33}$ 
ergs s$^{-1}$ (Seward et al.\ 2003). The spectrum is soft, 
but since absorption is comparable
in the two remnants, if this object were in Kes 27, we would expect 
40 counts ks$^{-1}$, twice as bright as the brightest observed
unresolved source.  The hardness ratio (defined in the notes to
Table~\ref{T:sources}) should be $\approx 0.5$, and no
variation or optical counterpart is expected.  As described below, 
we searched for such an object
and did not find one.  An upper limit of 0.8 counts ks$^{-1}$ yields
$L_x \leq 1.4 \times 10^{32}$ ergs s$^{-1}$ for any CCO in Kes 27.

\begin{center}
\begin{deluxetable*}{ccrcccccccc}
  \tabletypesize{\footnotesize}
  \tablecaption{{\sl Chandra} Source List \label{acis_source_list}}
\tablewidth{0pt}
  \tablehead{
  \colhead{Source} &
  \colhead{CXOU Name} &
  \colhead{Count rate $({\rm~cts~ks}^{-1})$} &
  \colhead{HR} &
  \colhead{$J$} &
  \colhead{$H$} &
  \colhead{$K$} &
  \colhead{$B$} &
  \colhead{$R$} &
  \colhead{$I$} &
  \colhead{Comments}\\
  \colhead{(1)} &
  \colhead{(2)} &
  \colhead{(3)} &
  \colhead{(4)} &
  \colhead{(5)} &
  \colhead{(6)} &
  \colhead{(7)} &
  \colhead{(8)} &
  \colhead{(9)} &
  \colhead{(10)} &
  \colhead{(11)} 
  }
  \startdata
   1 & J154722.45--534552.6 & 0.81$\pm$0.18 &  0.80$\pm$0.18
     & 14.07 & 13.54 & 13.45 & 17.42 & 16.05 & 15.27 & ? \\ 
   2 & J154737.93--534639.6 & 2.55$\pm$0.29 & -0.10$\pm$0.12
     & \nodata & \nodata & \nodata & \nodata & \nodata & \nodata
     & star?, varies \\ 
   3 & J154802.08--534549.3 & 1.53$\pm$0.23 & -0.19$\pm$0.16
     & 12.16 & 11.44 & 11.17 & 16.75 & 14.77 & 14.02
     & star \\ 
   4 & J154805.52--534050.5 & 0.75$\pm$0.17 &  -0.64$\pm$0.21
     & 11.50 & 11.23 & 11.17 & \nodata & \nodata & \nodata
     & star \\ 
   5 & J154805.67--534018.7 & 0.67$\pm$0.16 &  -0.44$\pm$0.25
     & 13.23 & 12.77 & 12.71 & 16.01 & 14.63 & 14.28 & star \\ 
   6 & J154808.11--534128.1 & 1.07$\pm$0.20 & -0.70$\pm$0.16
     & 10.84 & 10.67 & 10.61 & 11.69 & 11.59 & 11.56 & star \\ 
   7 & J154816.79--534125.5 & 7.03$\pm$0.46 &  0.95$\pm$0.03
     & \nodata & \nodata & \nodata & \nodata & \nodata & \nodata
     & AGN \\ 
   8 & J154817.90--535654.4 & 0.86$\pm$0.18 &  -0.44$\pm$0.21
     & 13.30 & 11.66 & 10.81 & 15.58 & 14.26 & 13.99 & star\\ 
   9 & J154819.85--534618.3 & 2.17$\pm$0.27 &  0.28$\pm$0.13
     & 10.83 & 9.74 & 9.31 & 19.49 & 15.63 & 13.95
     &  star \\ 
  10 & J154820.19--535447.9 & 1.05$\pm$0.20 & -0.54$\pm$0.18
     & 14.86 & 14.11 & 13.95 & 18.08 & 16.43 & 15.37 & star \\ 
  11 & J154820.78--534222.4 & 1.07$\pm$0.20 &  0.15$\pm$0.20
     & 14.01 & 13.04 & 12.65 & 20.61 & 18.54 & 17.11
     & star \\ 
  12 & J154822.10--535548.0 & 0.83$\pm$0.18 &  -0.29$\pm$0.22
     & 12.88 & 12.34 & 12.21 & 16.02 & 14.64 & 14.05 & star \\ 
  13 & J154826.38--534940.0 & 2.63$\pm$0.29 & -0.37$\pm$0.11
     & 13.42 & 12.70 & 12.47 & 17.56 & 14.87 & 14.28 & star \\ 
  14 & J154829.60--534627.1 & 3.57$\pm$0.34 & -0.53$\pm$0.09
     & 13.24 & 12.66 & 12.44 & \nodata & \nodata & \nodata
     & star, flare x10 \\ 
  15 & J154832.21--533736.3 & 1.80$\pm$0.25 &  0.64$\pm$0.12
     & \nodata & \nodata & \nodata & \nodata & \nodata & \nodata
     & AGN? \\ 
  16 & J154832.24--534028.0 & 0.97$\pm$0.19 &  0.89$\pm$0.14
     & \nodata & \nodata & \nodata & \nodata & \nodata & \nodata
     & AGN? \\ 
  17 & J154832.37--533916.8 &57.55$\pm$1.27 & -0.55$\pm$0.02
     & 10.24 & 9.86 & 9.75 & 12.20 & 11.28 & 10.90
     & star, varies x2  \\ 
  18 & J154832.96--535249.7 & 0.78$\pm$0.17 &  -0.52$\pm$0.22
   & \nodata & \nodata & \nodata & 20.02 & 18.70 & 16.68 & star, varies x3 \\ 
  19 & J154837.83--534955.7 & 2.47$\pm$0.29 & -0.52$\pm$0.11
     & 10.59 & 9.93 & 9.67 & 14.18 & 12.77 & 11.36 & star \\ 
  20 & J154839.05--534316.1 & 0.72$\pm$0.17 &  -0.19$\pm$0.25
     & 10.77 & 9.99 & 9.66 & 17.01 & 14.36 & 12.92 & star \\ 
  21 & J154839.32--534527.1 & 0.97$\pm$0.19 &  -0.28$\pm$0.20
     & 11.57 & 11.30 & 11.20 & 13.43 & 12.41 & 11.09 & star \\ 
  22 & J154840.21--534043.6 &13.20$\pm$0.62 & -0.71$\pm$0.0
     & 11.50 & 10.87 & 10.64 & \nodata & \nodata & \nodata
     & star, varies x3  \\ 
  23 & J154840.68--534635.9 & 0.67$\pm$0.16 &  -0.44$\pm$0.25
     & 12.52 & 12.16 & 12.08 & 14.57 & 14.06 & 13.69 & star \\ 
  24 & J154841.14--534005.2 & 0.78$\pm$0.17 &  -0.03$\pm$0.24
     & 13.74 & 13.22 & 13.01 & 17.27 & 15.55 & 14.91 & star \\ 
  25 & J154842.55--535610.1 & 0.83$\pm$0.18 &  0.81$\pm$0.18
     & 16.14 & 15.13 & 14.56 & \nodata & \nodata & \nodata
     & ? \\ 
  26 & J154842.77--535213.4 & 0.83$\pm$0.18 &  -0.36$\pm$0.22
     & 13.71 & 13.27 & 13.16 & 16.88 & 15.32 & 14.96 & star\\ 
  27 & J154853.11--534806.2 & 1.50$\pm$0.23 & -0.43$\pm$0.15 
     & 13.76 & 13.37 & 13.30 & 16.79 & 15.31 & 14.82 & star\\ 
  28 & J154915.48--534407.8 & 1.34$\pm$0.22 &  0.72$\pm$0.14
     & 15.91 & 14.97 & 14.34 & \nodata & \nodata & \nodata & ?\\ 
  29 & J154916.18--534244.9 & 1.34$\pm$0.22 &  0.80$\pm$0.13
     & \nodata & \nodata & \nodata & \nodata & \nodata & \nodata
     & AGN? \\ 
  30 & J154932.56--535026.4 & 0.59$\pm$0.16 &  0.00$\pm$0.29
     & 11.55 & 10.80 & 10.49 & 16.40 & 14.86 & 13.89 & star \\ 
\enddata
\tablecomments{
 Col.~(1): Generic source number. Col.~(2):
{\sl Chandra} X-Ray Observatory (unregistered) source name, following the
{\sl Chandra} naming convention and the IAU Recommendation for Nomenclature
(e.g., http://cdsweb.u-strasbg.fr/iau-spec.html).
Col.~(3): On-axis source broad-band count rate.
Col.~(4): The hardness ratio defined as
${\rm HR}=({\cal H}-{\cal S})/({\cal H}+{\cal S})$,
where ${\cal S}$ and ${\cal H}$ are the source count rates in the 0.3--1.5 and
1.5--7~keV bands, respectively.
This table only lists sources with individual signal-to-noise ratios greater
than 4 in the broad band (${\cal S}+{\cal H}$), with an exception made
for source 30, which has a S/N of 3.8 but is obviously discernible
in the tricolor image (Fig.3). 
Cols.~(5)--(7): $JHK$ magnitudes from the 2MASS catalog (Skrutskie et al.\ 2006).
Cols.~(8)--(10): $BRI$ magnitudes from the USNO-B1.0 catalog (Monet et al.\ 2003).
}
  \label{T:sources}
  \end{deluxetable*}
\end{center}

\begin{figure}[tbh!]
\centerline{ {\hfil\hfil
\psfig{figure=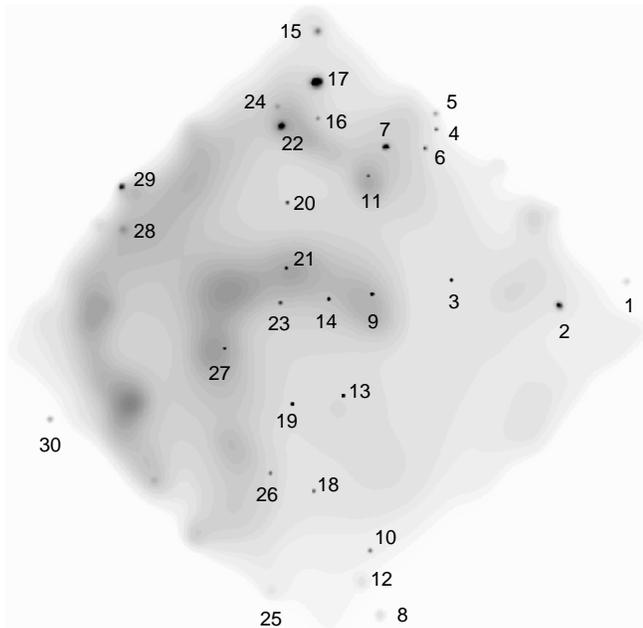,height=3.5in,angle=0, clip=}
\hfil\hfil}}
\caption{Identification of the unresolved sources listed in
Table~\ref{T:sources}, showing the adaptively smoothed 0.3--7~keV
emission in logarithmic gray scale.
}
\label{f:sources}
\end{figure}

We first searched for pointlike sources in the broad band
(0.3--7.0 keV) with a wavelet source detection algorithm.
Thirty sources with a signal-to-noise ratio S/N$\gsim 4$
in the field of view (the range of the four ACIS-I chips) are listed in
Table~\ref{T:sources} and identified in Figure~\ref{f:sources}. 
The spectrum of each source is characterized by a hardness ratio.
This is useful in identifying red stars and active galactic nuclei (AGNs). 
The count rates listed were taken from \Chandra\ automatic processing.
Optical data were taken from the Two Micron All Sky Survey (2MASS)
catalog (Skrutskie et al.\ 2006;
\footnote{See http://www.ipac.caltech.edu/2mass.} $JHK$) and
the USNO-B1.0 catalog (Monet et al.\ 2003;
\footnote{Available at http://www.nofs.navy.mil/data/fchpix.} $BRI$). 
If no cataloged source was within $1''$
of the X-ray position, no optical data are listed.  Most
Table~\ref{T:sources} sources have soft X-ray spectra and optical counterparts
with $0.0 < J-H < 0.8$ and $0.0 < H-K < 0.4$.
These are identified as stars and are probably of spectral type K or M
(Tokunaga 2000).  The
X-ray spectra of these clearly show less absorption than that of Kes 27, so
these are certainly foreground objects --- as well they should be.
Sources 9 and 11, with harder spectra, are probably highly reddened stars. 
  Even though not many photons were collected,
the X-ray flux from some sources is quite variable. The
approximate variation is listed. 

The bright source No.7 is quite hard, is strongly absorbed, and has no
counterpart.  This is probably a background AGN.  Sources 15, 16, and
29 are also hard with no counterparts and are likely to be AGNs.
The question marks in Table~1 for these sources indicate
classification based on hardness ratio.  There were not enough events
to determine a spectral shape.  A typical AGN with strength of 1 count
ks$^{-1}$ and with extinction corresponding to at least $2 \times
10^{22}\cm^{-2}$, the column density along the path to Kes~27, would have a hardness
ratio $\geq 0.7$ with $B \sim 27$, $R \sim 21$, and  $H \sim 13$~mag
(Seward 2000; NASA/IPAC Extragalactic Database
\footnote{See http://nedwww.ipac.caltech.edu.}).
There might be a counterpart in the infrared
($JHK$), but not in the optical bands ($BR$).
Sources 1,
25, and 28 are hard, with apparent counterparts.  Since these three
are far from the focal point, the X-ray location is not as accurate as
for the more central sources, and there is a higher probability that
 the counterparts listed are accidental.
Sources 25 and 28, with infrared counterparts only, could be AGNs.
Sources 2 and 18 are soft, with no 2MASS counterpart.  They are classified as
stars because they were observed to vary.

\begin{figure}[tbh!]
\centerline{ {\hfil\hfil
\psfig{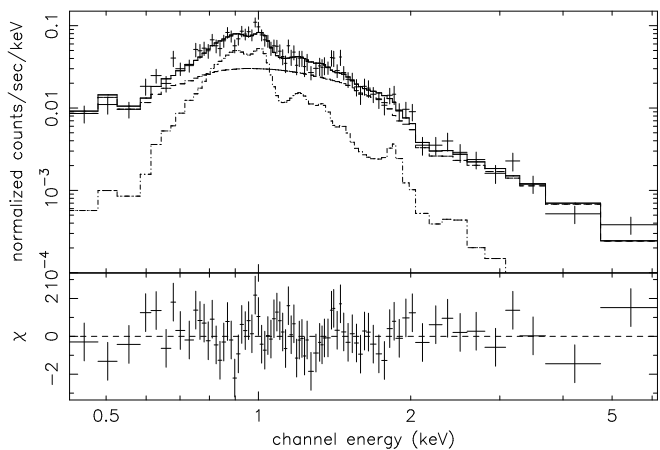}
\hfil\hfil}}
\caption{Spectrum of the northern hot spot (source 17), fitted with
a {\em vmekal}+{\em powerlaw} model. The source spectrum was extracted from
within a 3~$\sigma$ region, and an area of twice the 3~$\sigma$ size
surrounding the source region was used as the background.
The net count rate is $(5.63\pm0.12)\E{-2}\ps$. The spectrum
has been adaptively binned to achieve a
background-subtracted S/N of 4.
}
\label{f:nspot1}
\end{figure}

The brightest unresolved source, No.17, produced enough counts to derive
spectral parameters.  The spectrum
 can best be fitted with a thermal gas plus power law
model (see Fig.~\ref{f:nspot1}).
This source is weakly absorbed [$\NH\sim(6\pm4)\E{20}\cm^{-2}\ps$],
with power-law photon index $\Gamma\sim2.9\pm0.2$, gas temperature
$kT\sim0.77^{+0.07}_{-0.13}\keV$, probably overabundant Ne
([Ne/H]$\sim3.4^{+3.0}_{-2.3}$), and an unabsorbed 0.5--10~keV flux
$F_x\sim5.9\E{-13}\ergs\cm^{-2}\ps$ ($\chi^2=63.3$ with 73 degrees of freedom [dof]).
The spectrum of the second brightest unresolved source, No.22,
 can be fitted with a double thermal gas model
[$\NH\sim(6\pm1)\E{21}\cm^{-2}$, $kT_1\sim0.2^{+0.05}_{-0.02}\keV$,
$kT_2\sim1.3^{+1.4}_{-0.6}\keV$, $F_x\sim1.7\E{-12}\ergs\cm^{-2}\ps$, 
$\chi^2/{\rm dof}=14.3/17$].  These spectra and column densities
 are quite consistent with the identification as foreground stars.

Photons were sparse for most Table 1 sources.
A few sources were not easy to classify.
Source 2 is soft and has no optical counterpart, but it does not exhibit
enough interstellar absorption to be at the distance of the remnant.
Source 15
has no counterpart, and the hardness ratio is close to the 0.5 expected
from a CCO.  Since it is at the limb of the remnant, an AGN
identification is more likely.

Two interior sources in the previous \ROSAT\ image, Nos.\ 2 and 4,
were identified as possible PSR/PWN candidates (Seward et al.\ 1996).
These correspond respectively to \Chandra\ sources 14 and 19,
which are identified here as foreground stars.

\subsection{Spatial Analysis} \label{sec:spatial}

Figure~\ref{f:bband} presents the \Chandra\ 
image of diffuse emission in the $0.3--7.0\keV$ broad band.
Point sources were subtracted, and the pixel values in source regions were
replaced with values interpolated from the surrounding area.
The X-ray image was then exposure-corrected and 
adaptively smoothed to achieve a S/N of 4
(using the CIAO program {\em csmooth}).
Contours of the Molonglo Observatory Synthesis Telescope (MOST)
843 MHz radio emission observation of the remnant
(Whiteoak \& Green 1996) are superposed on the X-ray image.

\begin{figure}[tbh!] 
\centerline{ {\hfil\hfil
\psfig{figure=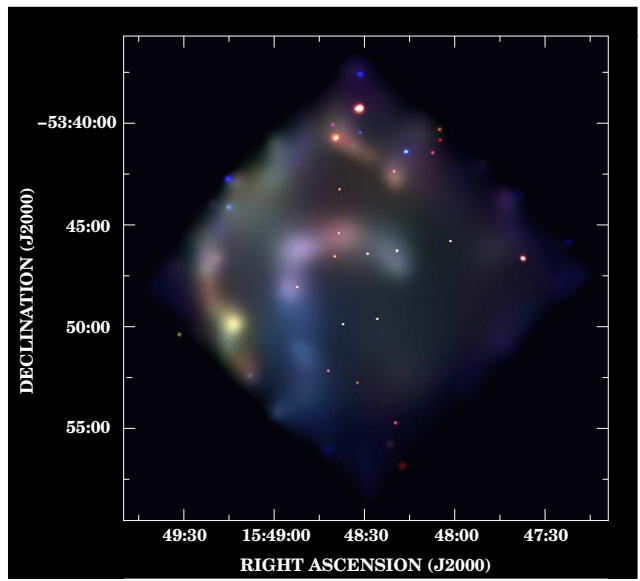,height=3.in,angle=0, clip=}
\hfil\hfil}}
\caption{
Tricolor ACIS-I image of \snr.  The X-ray intensities
in the 0.3--1.5, 1.5--2.2, and 2.2--7.0 keV bands are coded
red, green, and blue, respectively, and are scaled logarithmically
in the ranges 19--170, 19--290, and 27--270
photons$\cm^{-2}\ps\,\mbox{sr}^{-1}$.
The exposure-corrected X-ray maps were
adaptively smoothed to achieve a broadband (0.3--7 keV) S/N of 4
(using the CIAO program {\em csmooth}).
}
\label{f:3color}
\end{figure}

To demonstrate the energy dependence of the SNR morphology and
the point sources,
we also created a tricolor image in three energy bands,
as shown in Figure~\ref{f:3color}:
0.3--1.5 keV ({\em red}), 1.5--2.2 keV ({\em green}), and 2.2--7.0 keV
({\em blue}).
These three bands respectively contain the Mg, Si, and S lines,
which dominate the remnant's thermal emission (see \S\ref{sec:spec}),
and have comparable photon fluxes.
We first produced exposure maps in
the three bands and used these
for flat-fielding, accounting for bad-pixel removal,
correcting for telescope vignetting,
and correcting for variations in quantum efficiency across the detector.
The images in the three bands were then smoothed with an adaptive filter
(using program {\em csmooth}) to achieve a broadband (0.3--7 keV)
S/N of 4.

\begin{figure}[tbh!] 
\centerline{ {\hfil\hfil
\psfig{figure=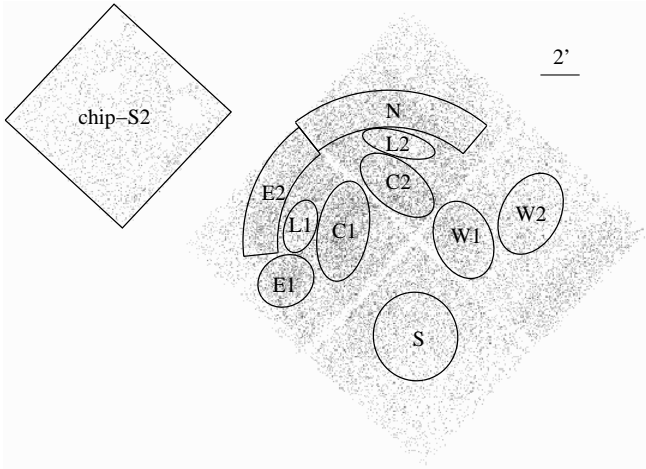,height=2.5in,angle=0, clip=}
\hfil\hfil}}
\caption{
Raw image of the ACIS observation (with point-like sources removed).
The labeled regions are used for spectrum extraction.
}
\label{f:reg}
\end{figure}

\begin{figure}[tbh!] 
\centerline{ {\hfil\hfil
\psfig{figure=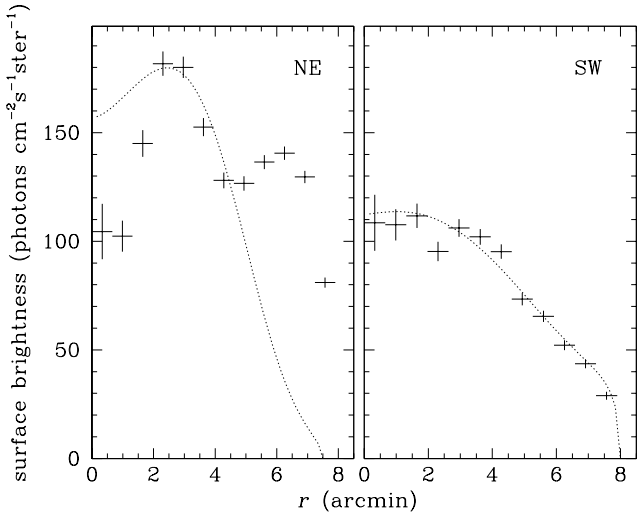,height=2.7in,angle=270, clip=}
\hfil\hfil}}
\caption{
Exposure-corrected radial X-ray (0.3--7 keV) surface profiles
for the northeast and southwest parts, centered at
R.A.=$\RA{15}{48}{37.9}$, decl.=$\decl{-53}{48}{01}.9$.
The NE part is from position angle $-46.8^{\circ}$ to $133.2^{\circ}$
and the SW part is from $133.2^{\circ}$ to $313.2^{\circ}$.
A blank-sky exposure on the ACIS-I chips has been used for background.
The dotted lines are plotted according to the White \& Long (1991) model,
relatively scaled to match the maximum brightness,
with parameters $(\tau, C)=(10,50)$ for the left panel and
$(\tau, C)=(20,56)$ for the right (see text).
}
\label{f:rp}
\end{figure}

Figures~\ref{f:bband} and \ref{f:3color} show that the X-ray--emitting
gas occupies a volume of linear size $\ge17'$ (the
size of the ACIS-I CCD chips)
and that the centroid of the X-ray emission lies in the eastern half of
the remnant.
The bright X-ray interior previously seen is now resolved into
two semicircular arcs in the northeastern half of the remnant.
The emission from the inner arc is harder than that from the outer arc
(see Fig.~\ref{f:3color}).
Apart from this bright, hard arc, there is an incomplete, soft arc
along the northeastern outskirt.
This outer arc is interior to the radio border.
The X-ray surface brightness peaks near the southeastern border,
where there is a bright X-ray knot (region E1 in Fig.~\ref{f:reg})
just inside the peak radio emission.
This phenomenon is very similar to that seen on the western rim
of SNR~3C~391, where the blast wave has been suggested to be propagating
into a small, dense region, causing drastic shock deceleration
or magnetic field compression and amplification (Chen et al.\ 2004).
The soft, bright, bar-like part of the outer arc in the north
appears coincident with a radio arc (Fig.~\ref{f:bband}).
The surface brightness radial profiles
are plotted in Figure~\ref{f:rp}.
The two peaks in the northeast profile
correspond to the two X-ray arcs, with mean radii of
$2.5'$ and $6'$. The southwest profile shows that the X-ray
emission fades with increasing radius, and no shell-like structure
can be discerned.

\subsection{Spectral Analysis of the Diffuse Emission}\label{sec:spec}

\begin{figure*}[tbh!] 
\centerline{ {\hfil\hfil
\psfig{figure=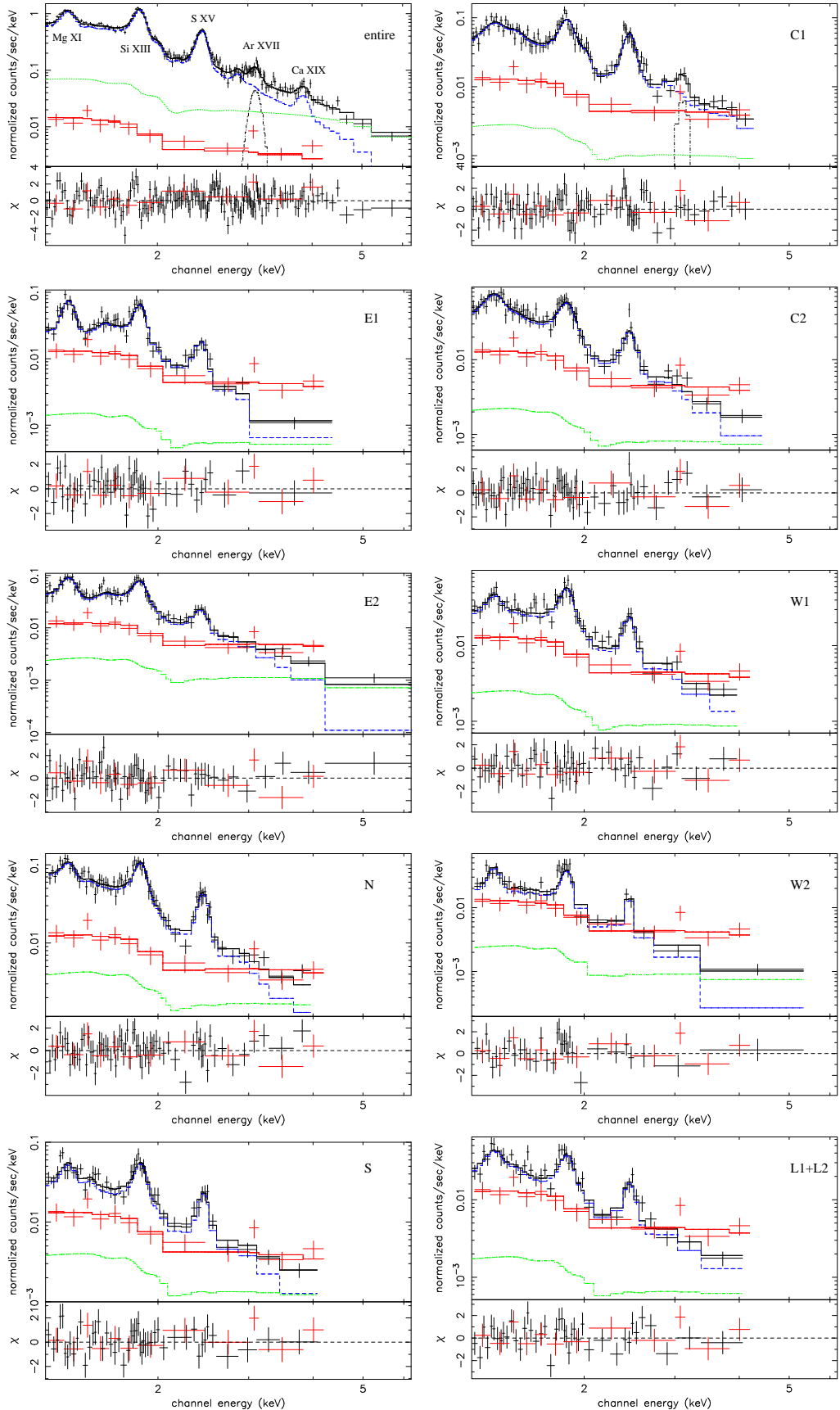,height=9in,angle=270, clip=}
\hfil\hfil}}
\caption{
ACIS spectra of the entire remnant and the small regions
shown in Fig.~\ref{f:reg}. The on-SNR spectra (black solid lines)
were rebinned to achieve a background-subtracted S/N of 4,
and the off-SNR one (red) was rebinned using a ratio of 3.
The net on-SNR spectra (blue) are fitted with an absorbed
{\em vpshock} model, except for the spectrum of the entire remnant,
for which an absorbed {\em vsedov} model is used.
The sky backgrounds (green) are scaled from the off-SNR
spectrum, which is mimicked by a {\em power-law} model.
The dot-dashed lines around
3.12 keV for the spectral fits of the entire remnant and region C1
are Gaussians accounting for the Ar~XVII line.
}
\label{f:spectra}
\end{figure*}

After removal of point sources, we extracted the overall
spectrum of the diffuse emission
from the whole area of the four ACIS-I chips (excluding
the peripheral edge)
and other on-SNR spectra from 10 substructures marked in
the raw-counts image shown in Figure~\ref{f:reg}.
Because the field of view of the four ACIS-I chips is
almost entirely covered by the remnant, a local background cannot
be obtained from these chips, and hence a double background
subtraction method was used.
We extracted an off-SNR source-free spectrum from the ACIS-S2 chip 
(excluding the peripheral edge again; Fig.~\ref{f:reg}).
The respective blank-sky background
\footnote{See http://cxc.harvard.edu/contrib/maxim/acisbg, period D.}
contributions estimated from the same regions
and scaled by the flux at 9.5--12~keV
\footnote{A high-energy band free of sky emission;
see also http://cxc.harvard.edu/contrib/maxim/acisbg.}
were subtracted from the on- and off-SNR spectra.
Individual on-SNR spectra were adaptively binned to achieve a
background-subtracted signal-to-noise ratio of 4 
and the off-SNR spectrum was binned using a ratio of 3.
Each on-SNR spectrum was jointly fitted together with the
off-SNR spectrum (Fig.~\ref{f:spectra}).
The on-SNR sky background was determined by scaling the off-SNR
emission according to the region sizes
(Fig.~\ref{f:spectra}, {\em green lines})
and is phenomenologically well described by a power law
with photon index $\sim0.7$.
Because of the poor signal-to-noise ratio of the off-SNR spectrum below 1.2~keV,
the counts below 1.2~keV in all the on-SNR spectra were ignored
so as to match the off-SNR one.
The {\small XSPEC} spectral fitting package (ver.\ 11.3)
\footnote{See http://heasarc.gsfc.nasa.gov/docs/software/lheasoft/xanadu/
xspec/xspec11/index.html.}
was used throughout.  For the foreground absorption, the cross sections
from Morrison \& McCammon (1983) were used, and solar abundances were assumed.

\subsubsection{Overall Spectral Properties} \label{sec:glo}
The spectrum of the entire remnant (Fig.~\ref{f:spectra}, top left)
shows distinct K$\alpha$ emission line features of metal elements:
Mg~XI ($\sim1.34\keV$),
Si~XIII ($\sim1.84\keV$),
S~XV ($\sim2.43\keV$),
Ar~XVII ($\sim3.12\keV$), and 
Ca~XIX ($\sim3.86\keV$),
confirming the thermal origin of the emission.
We fitted the net on-SNR spectrum with two absorbed non-equilibrium
ionization (NEI) thermal plasma models, {\em vpshock} and {\em vsedov}
(Borkowski et al.\ 2001; see Table~\ref{T:ent_spec}).
The former model characterizes the plasma parameters of a plane-parallel
shock, typified by a constant electron temperature and the shock
ionization timescale (the product of the electron density and the time
since passage of the shock). The latter is based on Sedov (1959)
dynamics,
typified by the mean and the electron temperatures immediately behind the
shock and by the ionization timescale (the product of the electron density
immediately behind the shock and the remnant's age).
Because the NEI models do not include emission from argon,
we added a Gaussian at 3.12~keV for the Ar line.
Since this spectrum comprises contributions from various regions
with different physical properties, it is not expected to be well
fitted by a model with a single thermal component.
However, it helps to sketch the overall properties
of the hot gas interior to the remnant.
The two models produce similar physical parameters, but
from comparison of the statistical goodness,
the {\em vsedov} model seems to better fit the spectrum than {\em vpshock}.
Both models produce elevated abundances
of S ($\sim1.9$--2.8 times solar) and Ca ($\sim$2.8--6.8),
indicating that the plasma is metal-enriched.
The postblast shock temperature $kT_s\sim0.37$--$0.51\keV$
derived with the {\em vsedov} model
corresponds to a mean gas temperature
$\approx1.27kT_s\sim0.47$--$0.65\keV$ for the Sedov (1959) case,
comparable to the mean temperature $\sim0.62$--$0.66\keV$
derived with the {\em vpshock} model.
The mean ionization timescale is $\lsim10^{12}\cm^{-3}\,\mbox{s}$,
which indicates that the remnant as a whole is close to,
but has not completely reached, ionization equilibrium.
The unabsorbed fluxes (0.5--$10\keV$) in the field of view
inferred from the two models are (2.3--$2.7)\E{-10}$
and (5.4--$8.6)\E{-10}\ergs\cm^{-2}\ps$, corresponding to X-ray luminosities
of (5.1--$6.0)\E{35}\du^2$ and (1.2--$1.9)\E{36}\du^2\ergs\ps$,
respectively.

\begin{center}
\begin{deluxetable*}{c|cc}
\tabletypesize{\footnotesize}
\tablecaption{Spectral Fitting Results for the Entire Remnant
  and Estimates of the Das Density}
\tablewidth{0pt}
\tablehead{
\colhead{Parameters} \vline & \colhead{\em vpshock} & \colhead{\em vsedov}
}
\startdata
$\chi_{\nu}^{2}$ (dof) & 1.89 (162) & 1.57 (161) \\
$\NH$ ($10^{22}\cm^{-2}$) & $2.3\pm0.1$ & $2.3\pm0.1$ \\
Temperature ($\keV$) & $kT_x=0.63^{+0.03}_{-0.01}$$^{\,\,\rm a}$
  & $kT_s=0.41^{+0.10}_{-0.04}$$^{\,\,\rm b}$ \\
       &       & $kT_e=0.41^{+0.04}_{-0.03}$$^{\,\,\rm b}$ \\
Ion.\ Timescale ($10^{11}\cm^{-3}\,{\rm s}$)$^{\rm c}$ & $12.0^{+2.7}_{-4.8}$
  & $4.19^{+1.15}_{-0.83}$ \\
$f\nel\nH V/\du^{2}$ ($10^{58}\cm^{-3}$)$^{\rm d}$
  & $2.01^{+0.10}_{-0.19}$ & $3.36^{+0.85}_{-0.72}$ \\
{[Mg/H]} & $0.85^{+0.09}_{-0.08}$ & $0.63^{+0.08}_{-0.07}$ \\
{[Si/H]} & $1.00^{+0.07}_{-0.06}$ & $1.13^{+0.12}_{-0.09}$ \\
{[S/H]} & $2.10^{+0.20}_{-0.22}$ & $2.47^{+0.36}_{-0.28}$ \\
{[Ca/H]} & $4.87^{+1.92}_{-1.77}$ & $4.45^{+1.84}_{-1.66}$ \\
Flux ($10^{-10}\ergs\cm^{-2}\ps$)
  & $2.56^{+0.12}_{-0.25}$ & $6.88^{+1.74}_{-1.47}$ \\
$\nH/f^{-1/2}\du^{-1/2}$ ($\cm^{-3}$)\hspace{1mm}$^{\rm e}$
  & $0.34^{+0.01}_{-0.02}$ & $0.44^{+0.06}_{-0.05}$
\tablecomments{The unabsorbed fluxes are in the 0.5--$10\keV$ band.
   The net count rates of the on- and off-source spectra are
   $(8.52\pm0.06)\E{-1}$ and $(1.86\pm0.14)\E{-2}$ cts$\ps$,
   respectively. Confidence ranges are at the 90\% level.
}
\enddata
  \tablenotetext{a}{\phantom{0} Mean electron temperature.}
  \tablenotetext{b}{\phantom{0} Mean shock temperature and electron
       temperature immediately behind the shock front, respectively.}
  \tablenotetext{c}{\phantom{0} The ionization timescale is defined as
       the product of the electron density and the time since the passage
       of the shock in the {\em vpshock} model, and as the product of
       the electron density immediately behind the shock and the remnant's
       age in the {\em vsedov} model. The value listed for {\em vpshock}
       is the upper limit for the range of the timescale.}
  \tablenotetext{d}{\phantom{0} Where $f$ denotes the filling factor of the hot gas.}
  \tablenotetext{e}{\phantom{0} In the estimate of the densities, we assume
       a spherical volume of diameter $17'$ for the X-ray--emitting gas.}
\label{T:ent_spec}
\end{deluxetable*}
\end{center}

\begin{figure}[tbh!] 
\centerline{ {\hfil\hfil
\psfig{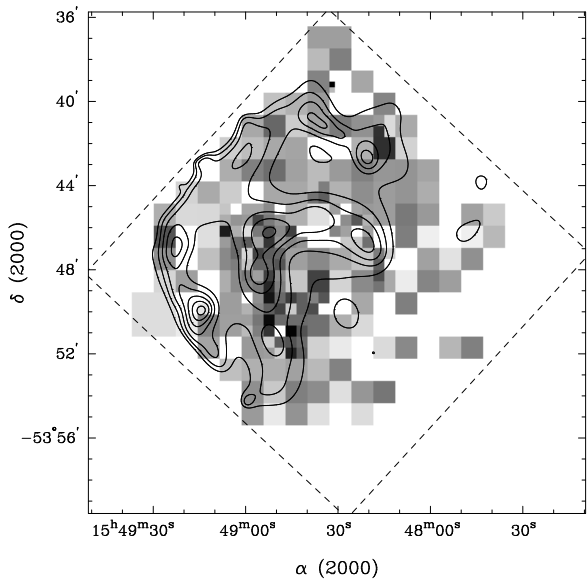}
\hfil\hfil}}
\caption{
Equivalent width image of the S~He$\alpha$ ($\sim2.46\keV$) line
on a square root scale
overlaid with the broadband (0.3--7 keV) diffuse X-ray contours (at 
seven logarithmic scale levels between
120 and 400 photons$\cm^{-2}\ps\,\mbox{sr}^{-1}$).
The energy range used to extract the S~line image is 2.28--2.70 keV,
while the low- and high-energy ranges around
the S~line used to estimate the underlying continuum
are 2.07--2.28 and 3.30--3.70 keV.
The dashed box denotes the field of view of ACIS-I.
}
\label{f:ew}
\end{figure}

The spatial distribution of the relative strength of the
S He$\alpha$ emission is shown in
the equivalent width (EW) image of this line in Figure~\ref{f:ew}.
This image was constructed using a method similar to those of
Hwang et al.\ (2000) and Park et al.\ (2002).
A major difference of our method from theirs
is that we rebin the data using an adaptive mesh, with each
bin in each narrow-band image including at least 10 counts
(see Warren et al.\ 2003 for similar binning).
The image shows that the S~line emission is distributed over
a broad region but is strongest in the bright eastern portion of the
inner arc.
The Ca line is very weak, and we did not produce a Ca EW image.

\subsubsection{Small-Scale Properties}
The spectra of the 10 substructures
all show distinct emission lines,
indicating optically thin thermal plasma.
They were fitted with the {\em vpshock} model, and
the spectral fit parameters are summarized in Table~\ref{T:xspec}.
A Gaussian at $\sim3.1\keV$ is added to account for the
Ar~XVII line in the region C1 spectrum.
The model fits
allow the abundances of key line-producing elements to vary while leaving
other abundances fixed to the solar value.
Estimates of the hydrogen number densities ($\nH$)
inferred from the volume emission measures
($fn_e\nH V$, where $f$ is the filling factor of the hot gas
and $n_e\sim1.2\nH$ is assumed)
are also given in Table~\ref{T:xspec}
(and Table~\ref{T:ent_spec} for the entire remnant).
In the derivation of the densities we have assumed that
the three-dimentional shapes of the elliptical regions
(E1, C1, C2, W1, W2, S, L1, and L2) are ellipsoids, those of the
arclike regions (E2 and N) are projections of spherical segments,
and that of the entire X-ray--emitting volume is a sphere of
diameter $17'$.
Deviations from these assumptions and the nonuniformity of
the X-ray--emitting plasma are consolidated into 
the factors $f$ of the individual regions.
The regions show significant spectral variation.
The gas temperatures ($\sim 1\keV$) along the inner semicircular arc
(regions C1 and C2) are higher than those ($\sim0.4$--$0.7\keV$)
along the softer outer X-ray arc (regions E1, E2, and N).
The densest gas in the eastern region E1 is consistent with the suggestion,
based on the radio and X-ray brightness peaks, that the forward shock
encounters a small dense region there (\S~\ref{sec:spatial}).
The gas density of the low surface brightness regions (L1 and L2)
between the two arcs is lower than that in the two bright arcs.
The ionization timescales for the small-scale regions
(except E1 and E2) are around $10^{12}\cm^{-3}\mbox{s}$ or
even higher, and so they may be close to ionization equilibrium,
but the ionization timescales for regions E1 and E2 (along
the eastern outer shell) are relatively low
[(0.6--$3.0)\E{11}\cm^{-3}\mbox{s}$],
indicating that the plasma is far from ionization equilibrium.
The sulfur abundance in most of the small-scale regions aside from E2
is somewhat higher than solar and seems to peak in region C1, the eastern
portion of the inner arc, as shown in the S He$\alpha$ line distribution
in the EW map.

\begin{center}
\begin{deluxetable*}{cc|ccccccccc}
\tabletypesize{\footnotesize}
\tablecaption{{\em vpshock} Fitting Results
  and Estimates of the Gas Density}
\tablewidth{0pt}
\tablehead{
\colhead{Regions} & \colhead{Net Count Rate} \vline &
\colhead{$\chi^{2}_{\nu}$ (d.o.f.)} & \colhead{$\NH$} &
\colhead{$kT_{x}$} & \colhead{$n_e t$} &
\colhead{$f\nel\nH V/\du^{2}$\hspace{1mm}$^{\rm a}$} &
\colhead{[Mg/H]} & \colhead{[Si/H]} & \colhead{[S/H]} & 
 $\nH/f^{-1/2}\du^{-1/2}$\hspace{1mm}$^{\rm b}$\\
\noalign{\smallskip}
\colhead{} & \colhead{($10^{-2}$ cts$\ps$)} \vline & \colhead{} &
\colhead{($10^{22}\cm^{-2}$)} &
\colhead{(keV)} & \colhead{($10^{11}\cm^{-3}\,{\rm s}$)} &
\colhead{($10^{56}\cm^{-3}$)} & \colhead{} &
\colhead{} & \colhead{} & \colhead{(cm$^{-3}$)}
}
\startdata
E1 & $4.00\pm0.11$ & 1.07 (54) & $2.7\pm0.2$ &
  $0.47\pm0.04$ & $1.68^{+1.36}_{-0.42}$ & $36.2^{+14.3}_{-11.1}$ &
  $0.74^{+0.28}_{-0.20}$ & $0.86^{+0.25}_{-0.19}$ & $1.86^{+0.82}_{-0.63}$ &
  $2.1^{+0.4}_{-0.3}$\\
E2 & $6.05\pm0.14$ & 0.92 (69) & $2.5^{+0.1}_{-0.2}$ &
  $0.60^{+0.07}_{-0.06}$ & $0.89^{+0.76}_{-0.34}$ & $29.1^{+10.2}_{-4.6}$ &
  $0.68^{+0.22}_{-0.17}$ & $0.74^{+0.21}_{-0.08}$ & $1.29^{+0.52}_{-0.45}$ &
  $0.73^{+0.13}_{-0.06}$\\
N & $7.58\pm0.15$ & 1.02 (75) & $1.9\pm0.2$ &
  $0.60^{+0.04}_{-0.02}$ & 18.9 ($>5.6$) & $16.7^{+3.4}_{-4.2}$ &
  $0.67^{+0.14}_{-0.21}$ & $1.06^{+0.27}_{-0.22}$ & $2.48^{+0.94}_{-0.66}$ &
  $0.43^{+0.04}_{-0.05}$\\
S  & $4.18\pm0.12$ & 0.98 (54) & $2.3\pm0.2$ &
  $0.72^{+0.07}_{-0.05}$ & 12.7 ($>3.8$) & $7.2^{+1.9}_{-1.0}$ &
  $1^{\rm c}$ & $1^{\rm c}$ & $1.73^{+0.73}_{-0.55}$ &
  $0.25^{+0.03}_{-0.02}$\\
C1  & $8.19\pm0.16$ & 1.26 (90) & $1.8\pm0.6$ &
  $1.21^{+0.45}_{-0.39}$ & $3.7^{+22.6}_{-1.6}$ & $4.7^{+6.2}_{-1.3}$ &
  $0.83^{+0.52}_{-0.38}$ & $1.44^{+0.43}_{-0.32}$ & $2.97^{+0.39}_{-0.55}$ &
  $0.43^{+0.28}_{-0.06}$\\
C2  & $5.13\pm0.13$ & 0.85 (68) & $2.1^{+0.2}_{-0.3}$ &
  $0.76^{+0.25}_{-0.09}$ & $7.2^{+31.8}_{-5.0}$ & $6.7^{+2.4}_{-2.0}$ &
  $1.22^{+0.69}_{-0.45}$ & $1.10^{+0.40}_{-0.28}$ & $1.92^{+0.88}_{-0.59}$ &
  $0.63^{+0.11}_{-0.09}$\\
W1  & $4.12\pm0.11$ & 1.00 (54) & $2.6^{+0.3}_{-0.2}$ &
  $0.69^{+0.05}_{-0.04}$ & $7.6^{+72.2}_{-5.2}$ & $9.8\pm2.7$ &
  $0.84^{+0.47}_{-0.40}$ & $1.05^{+0.38}_{-0.29}$ & $1.79^{+0.78}_{-0.56}$ &
  $0.37\pm0.05$\\
W2  & $2.73\pm0.10$ & 1.15 (39) & $2.3^{+0.3}_{-0.4}$ &
  $0.62^{+0.05}_{-0.07}$ & 7.3 ($>2.0$) & $6.7^{+3.8}_{-2.0}$ &
  $1^{\rm c}$ & $1^{\rm c}$ & $1.92^{+1.15}_{-0.81}$ &
  $0.30^{+0.09}_{-0.04}$\\
L1+L2 & $3.21\pm0.10$ & 1.17 (47) & $2.2^{+0.2}_{-0.3}$ &
  $0.89\pm0.06$ & $5.2^{+13.6}_{-2.8}$ & $3.9\pm1.3$ &
  $1^{\rm c}$ & $1^{\rm c}$ & $1.60^{+0.60}_{-0.54}$ &
  $0.26\pm0.04$
\tablecomments{
   The net count rates of the on-source spectra are listed in the second
   column; that of the off-source spectrum is $(1.86\pm0.14)\E{-2}$ cts$\ps$.
   Confidence ranges are at the 90\% level.
}
\enddata
  \tablenotetext{a}{\phantom{0} Where $f$ denotes the filling factor of the hot gas.}
  \tablenotetext{b}{\phantom{0} In estimating the densities, we assume
    oblate spheroids for elliptical regions E1 (with half-axes
    $1.50'\times1.50'\times1.29'$), C1 ($2.62'\times2.62'\times1.31'$),
    and C2 ($2.22'\times2.22'\times1.18'$),
    ellipsoids (with half line-of-sight size to be the outer radius of
    shell region N of $8'$) for regions W1 (with half-axes
    $2.11'\times1.49'\times8'$), W2 ($2.22'\times1.50'\times8'$),
    S ($2.34'\times2.20'\times8'$), L1 ($1.43'\times0.82'\times8'$),
    and L2 ($1.87'\times0.84'\times8'$), and spherical segments for regions
    E2 (with inner and outer radii $5.8'$ and $7.6'$ and sector angle
    $80^{\circ}$) and N ($5.8'$, $8.0'$, $85^{\circ}$).}
  \tablenotetext{c}{\phantom{0} Fixed to the solar abundance.}
\label{T:xspec}
\end{deluxetable*}
\end{center}

\section{Discussion}
\snr\ has been classified as a thermal composite (or
mixed-morphology) remnant. 
It is unusual in that
the diffuse thermal X-ray emission peaks not in the center,
but along two concentric semicircular arcs.
The centroid of the X-ray emission lies in the eastern half of
the remnant, and the emission of the inner arc is harder than 
that of the outer region.

The thermal properties of \snr, as inferred from \S~\ref{sec:spec},
suggest that the X-ray--emitting gas is essentially close to equilibrium 
of ionization (except along the eastern rim)
and enriched with sulfur and calcium.
If we adopt $\nH\sim0.4f^{-1/2}\du^{-1/2}\cm^{-3}$ as
the mean gas density,
then the mass of hot gas is $M=1.4\nH\mH fV\sim110f^{1/2}\du^{5/2}\Du^3\Msun$,
where $\mH$ is the hydrogen atomic mass and
$\Du$ is the remnant's diameter $D$ scaled by the radio size,
$\sim20'$ (thus $D\approx25\Du\du\parsec$).
The thermal energy contained in the remnant is
$E_{th}=(3/2)2.3\nH kT fV\sim3\E{50}f^{1/2}\du^{5/2}\Du^3\ergs$,
where a mean temperature $kT\sim0.6\keV$ for the hot gas is used.
With a shock temperature $kT_s\sim0.4\keV$,
the shock velocity would be
$v_s=(16kT_s/3\bar{\mu}\mH)^{1/2}\sim580\km\ps$,
where the mean atomic weight $\bar{\mu}=0.61$.
The dynamical age of the remnant,
$\sim D/5v_s\sim8.4\E{3}\du\Du\yr$, is derived
with the Sedov (1959) solution, which,
because of the unusual morphology and complicated gas components,
is only an approximation.
The age estimate ($\sim3.5\E{3}\yr$) based on the \ROSAT\ X-ray
observation used a gas temperature ($3.2\keV$) that was too high
(Seward et al.\ 1996).
The Sedov model fit yields an ionization timescale
$\tau_{\rm Sedov}\sim4.2\E{11}\cm^{-3}\,\mbox{s}$,
which implies a dynamical age
$t\approx\tau_{\rm Sedov}/(1.2\times4\nH)\sim7\E{3}f^{1/2}\du^{1/2}\yr$
(where we have assumed that the preshock density is similar to
the mean interior density $\nH$).
This is similar to the above Sedov age.
On the other hand, the upper limit on
the mean ionization timescale derived with the {\em vpshock} model
(Table~\ref{T:ent_spec}) gives an age
$\sim9\E{4}f^{1/2}\du^{1/2}\yr$.
This is similar to the ionization age
($>8\E{4}\yr$) found in the \ASCA\ X-ray study (Enoguchi et al.\ 2002)
and is comparable to the remnant's radiative cooling
timescale, $t_{\rm cool}\sim4\E{4}E_{51}^{0.24}[n_0/(0.4\cm^{-3})]^{-0.52}\yr$,
where $E_{51}$ is the SNR's explosion energy in units of $10^{51}\ergs$
and $n_0$ is the preshock density (Falle 1981).
In this case the postshock temperature would be $\lsim1\E{6}\K$
and thus the gas would be X-ray--faint.
This does not agree with the postshock temperatures derived from
the spectral fits and the fact that the eastern
rim of the SNR is X-ray--bright.
A possibility to reconcile the different estimates might be for
the X-ray emitting gas to have a very low filling factor,
$f\lsim1\E{-2}\Du^2\du$, but this seems unlikely.


The mass of the X-ray--emitting gas, $\sim 10^2 f^{1/2} \Msun$,
suggests that the SNR is not in the free-expansion phase and is
probably dominated by the swept-up ambient medium.  Even so, 
Kes~27 appears to be enriched in S and Ca, and this is 
not a unique case. One sample of
23 thermal composite (or mixed-morphology) SNRs
showed that 10 of them to be detected as metal-enhanced (Lazendic \& Slane 2006).
The contribution of supernova ejecta has been suggested to be the cause of
both the elevated metal abundances and the X-ray brightening in the
central regions. For example, the overabundant Mg, Si, and (possibly) S
in thermal composite SNR~G290.1$-$0.8 (Slane et al.\ 2002)
and the Ne, Si, and S enhancement in W44 are ascribed to supernova ejecta
(Shelton et al.\ 2004).

Environments of nonuniform density are often seen around SNRs, especially
those adjacent to dense clouds.
This sometimes is consistent with systematic variation of the intervening
hydrogen column density along the line of sight to the SNR, as for
3C~391 (Chen et al.\ 2004) and 3C~397 (Safi-Harb et al.\ 2005).
\snr's HI density enhancement is found to the east of the remnant
(McClure-Griffiths et al.\ 2001), and
the Spitzer IRAC 8~$\mu$m observation (Reach et al.\ 2006) shows
a large-scale environment with a density gradient
increasing from west to east.
However, our X-ray study of \snr\ does not show a clear trend of
$\NH$ (and spectral) variations along the density gradient
(although E1 is the only region showing a significantly higher
value; Table~\ref{T:xspec}).
This may indicate that the environmental hydrogen column density
may take only a very low fraction in the intervening interstellar column
($\sim2\E{22}\cm^{-2}$), either because the cloud complex in the vicinity
is not very compact or bacause it is not very large.

The double X-ray rings are not expected in either the cloud
evaporation or the thermal conduction model.
In the latter (Cox et al.\ 1999; Shelton et al.\ 1999),
thermal conduction in the remnant smooths out the temperature gradient
from the hot interior to the cooler shell and increases the central density,
resulting in luminous X-rays in the interior; this effect is
dominant in the radiative stage.
This model predicts a centrally peaked X-ray surface brightness
for the SNR
and, usually, a high ionization timescale for the hot gas,
inconsistent with our observation.
The cloud evaporation model (White \& Long 1991) suggests that
when an SNR expands in an inhomogeneous interstellar medium 
whose mass is mostly contained in small clouds,
the clouds engulfed by the blast wave can be evaporated
to slowly increase the density of the interior hot gas;
as a result, the SNR appears internally X-ray--brightened.
This model can reproduce an X-ray--bright inner ring like that
in the northeastern half of \snr\ and can also reproduce a fading-out radial
surface brightness profile like that in the southwestern half
(see Fig.`\ref{f:rp}).
A representative fit to the inner ring's brightness profile using the
cloud evaporation model requires the ratio of the cloud evaporation
timescale to the SNR's age to be $\tau=10$ and the ratio of the mass of
the cloudlets to the mass of the intercloud medium to be $C=50$.
The fit is not unique, and combinations of $\tau=8$--12 and
$C=40$--60 produce similar profiles (relatively scaled to match
the maximum brightness).
A representative fit to the southwest radial brightness profile would need
$\tau=20$ (15--25) and $C=56$ (45--70).
However, the generally higher inner gas temperature as compared with the outer
region is not expected for such high $\tau$- and $C$-values
in the evaporation model. 
Moreover, this model overestimates the surface brightness inside the
inner ring and cannot reproduce the outer X-ray ring in the northeast.
It also cannot naturally explain why the physical conditions exemplified
by the parameters $\tau$ and $C$ should be different in the two halves
and why there are not two rings in the southwestern half, in contrast
to the northeastern.

\begin{figure}[tbh!] 
\centerline{ {\hfil\hfil
\psfig{figure=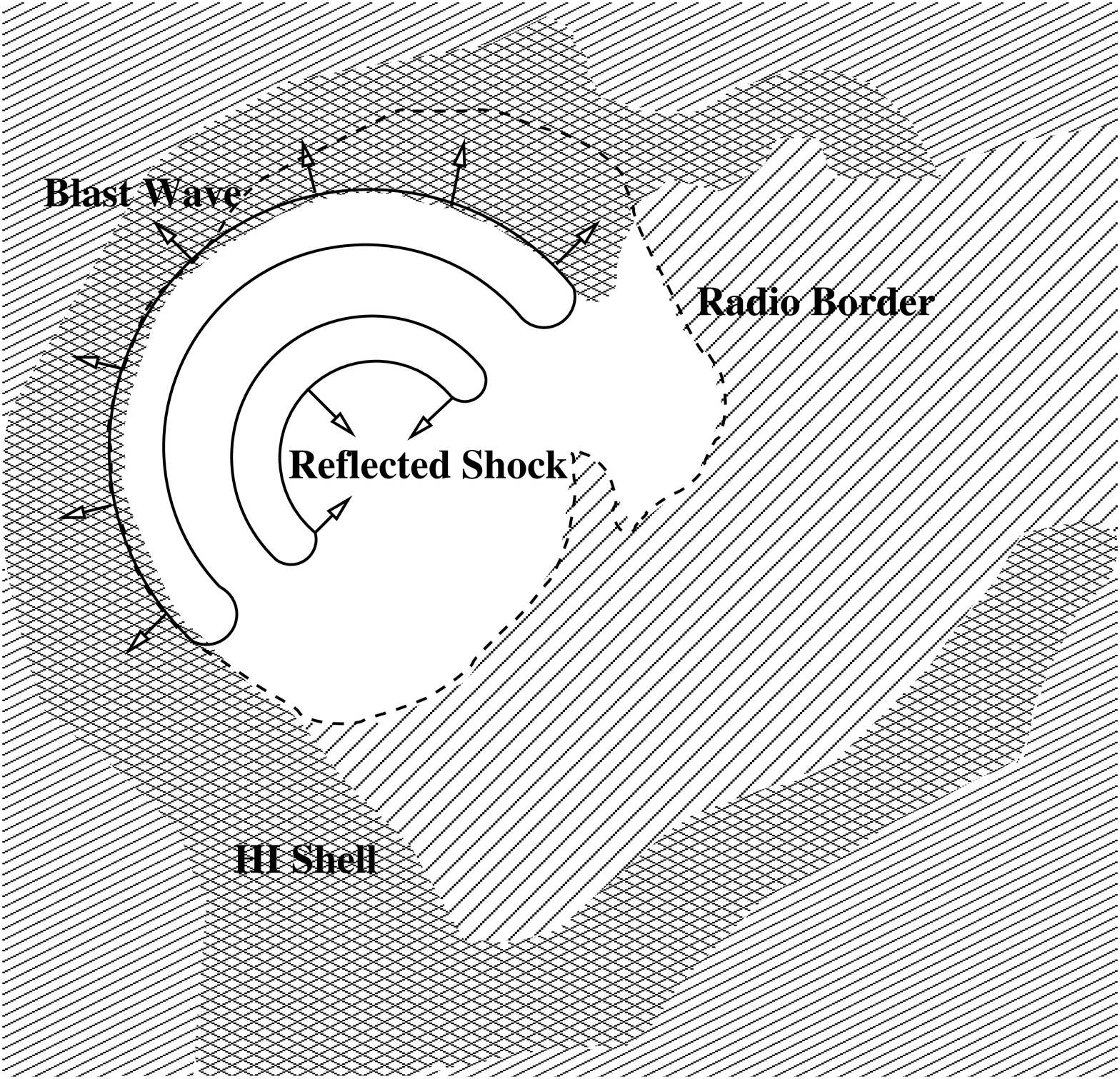,height=3.5in,angle=0, clip=}
\hfil\hfil}}
\caption{
Diagram showing the structure and environment of the Kes~27 remnant.
}
\label{f:diagram}
\end{figure}

Another scenario is required to account for
the unusual X-ray morphology of \snr,
which is different from that of other thermal composite remnants.
Actually, the explanation must be consistent with the following facts:
(1) The large-scale environment around \snr\ has a density gradient
increasing from west to east, as revealed in the infrared (Reach et al.\ 2006).
(2) As sketched in Figure~\ref{f:diagram}, the remnant is surrounded by
 a ``crocodile's mouth''--like \HI\ shell (McClure-Griffiths et al.\ 2001).
(3) In X-rays, the remnant is brightened in the northeastern half and 
 fades out to the southwestern low-density region.
(4) The inner and outer X-ray arcs are both located in the northeastern half,
 where the ambient medium is denser than in the southwest.
(5) The outer arc is in contact with the \HI\ shell.
(6) The inner arc is hotter than the outer arc.
(7) As noted by Enoguchi et al.\ (2002), the radio polarization
observed by Milne et al.\ (1989) is strong in both the eastern rim
and the center.

We suggest that both the outer and inner arcs represent shock waves,
propagating outward and inward, respectively.
The observed polarization in the eastern rim and the center is due to
shocked and compressed matter in the two arcs.
The present inward shock is not the initial reverse shock caused by
expansion into low-density circumstellar material. The remnant is
middle-aged, and this has long passed (McKee 1974). The present inward
shock is a reverse shock due to the distant HI shell.


A comprehensive scenario is as follows:
The massive progenitor exploded in a cavity excavated by
the strong stellar wind and ionizing radiation.
Because of the density gradient in the environmental medium,
the wind cavity had a small radius in the east and a large radius
in the west, or may even not have been confined in the west.
After the supernova explosion, in the west and southwest, the blast wave
propagated into a low-density medium, and the X-ray emission is faint.
In the north and east, the blast wave has hit
the cavity wall.  A transmitted shock is now propagating into the
dense wall, and a reflected shock is traveling back toward the remnant's center,
shocking the supernova ejecta to higher temperature.
The outer arc contains gas heated by the transmitted shock,
and the inner arc is the interior metal-enriched gas heated by
the reflected shock (see again Fig.~\ref{f:diagram}).
Shock-heated S atoms in the compressed inner arc
emit a strong He$\alpha$ line.

\begin{figure}[tbh!]
\centerline{ {\hfil\hfil
\psfig{figure=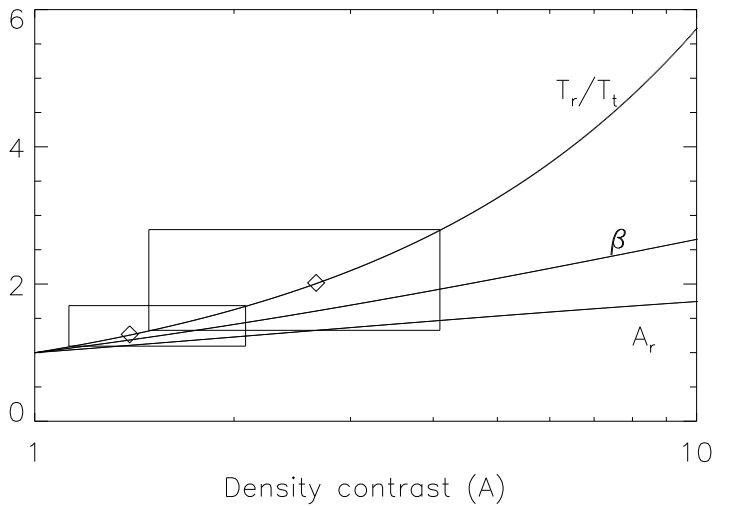,height=3.in,angle=0, clip=}
\hfil\hfil}}
\caption{Parameters for the reflected shock. The large box
 represents the value ranges for region pair E2 and C1, and the
 small box is for region pair N and C2 (see text).}
\label{f:refl_shock}
\end{figure}

Chen et al.\ (2003) considered a thin-shell model for a blast shock
hitting a wind-cavity wall but did not include shock reflections.
Dwarkadas (2005, 2007) simulated shocks reflected back from a cavity
wall, and a picture of shock reflection and transmission similar to
what is seen in \snr\ was predicted. To capture the essential physics
essence, we use the theory of reflected shocks described by Sgro (1975)
to estimate some relevant parameters.
For a shock reflected by a dense wall, the temperature ratio between 
the post--reflected-shock gas and post--transmitted-shock gas is given by
\begin{equation}
\frac{T_r}{T_t}=\frac{A}{A_r}, \label{eq:Tratio}
\end{equation}
where $A_r$ is the density contrast between the post--reflected-shock
gas and
post--incident-shock gas and $A$ is the wall-to-cavity
density contrast, which is related to $A_r$ as
\begin{equation}
A=\frac{3A_r(4A_r-1)}{\{[3A_r(4-A_r)]^{1/2}-\sqrt{5}(A_r-1)\}^2}.
\end{equation}
The pressure ratio of the post--transmitted-shock gas to the
post--incident-shock gas is
\begin{equation}
\beta=\frac{4A_r-1}{4-A_r}.
\end{equation}
Using these equations, we plot the functional relations of
$T_r/T_t$, $A_r$, and $\beta$ with $A$ in Figure~\ref{f:refl_shock}.
Regions E2 versus C1 and N versus C2 represent two pairs of
transmitted versus reflected shocks.
The ranges of the temperature ratios obtained from Table~\ref{T:xspec}
are shown in Figure~\ref{f:refl_shock}.
Because there is a large scatter in the fitted temperatures, there are
large uncertainties in the wall-to-cavity density $A$, which seems
to be $\sim$ 2.
The density contrast between the post--reflected-shock gas and 
postincident shock gas $A_r$ is around 1.3.
If regions W1, W2, and S can be taken as roughly indicative of
a post--incident-shock gas
density $\sim(0.2-0.4)f^{1/2}\du^{1/2}\cm^{-3}$ (Table~\ref{T:xspec}), then 
the post--reflected shock density $0.4-0.6f^{1/2}\du^{1/2}\cm^{-3}$
in C1 and C2 is consistent with the values of $A_r$ (although this estimate
is crude, because the filling factor $f$ may vary from region to region).

Such a reflected shock launched from the cavity wall is
consistent with the observed properties of \snr.
As the remnant continues to age, 
the transmitted shock in the dense medium will become fainter in X-rays,
while the reflected shock will converge in the center, producing a truly
centrally X-ray--brightened morphology.
The reflection of shocks is thus another mechanism that may produce
 thermal composite remnants.

\section{Summary}
We have observed the thermal composite SNR \snr\ with the ACIS-I
detectors on \Chandra\ and performed
a spatially resolved spectroscopic study of the remnant.
The main results are summarized as follows.
\begin{enumerate}
\item We have detected 30 pointlike X-ray sources superposed on the
remnant, with S/N$\gsim4$. Most are foreground stars, and some are probably
background AGNs.  None has the properties expected from
 a central compact stellar remnant. The upper limit to the luminosity of
 a CCO or a diffuse PWN is $\sim 1.5 \times 10^{32}$ ergs~s$^{-1}$.
\item The X-ray spectrum of \snr\ is characterized by K~lines
from ionized Mg, Si, S, Ar, and Ca.
Most of the X-ray--emitting regions are found to be enriched in sulfur.
Calcium is also over-abundant in the remnant.
The hot gas in the remnant is essentially close to ionization
equilibrium (except for that along the eastern rim).
\item Images of the remnant show previously unseen double X-ray
arc structures.  There is an outer X-ray arc and an inner,
harder, concentric arc. These two shell-like features are located 
in the northeastern
half of the remnant, and the X-ray surface brightness fades away with
increasing radius to the southwest. The X-ray intensity peak coincides with
the radio-bright region along the eastern border.
The gas in the inner region is at higher temperature.
\item
The overall morphology can be explained by the evolution of the
remnant in an ambient medium with a density gradient increasing
from west to east.
The remnant was probably born in a preexisting cavity
created by the progenitor star.  The outer shock has hit the cavity
wall and is now in the denser wall material, and the inner arc
indicates gas heated by a reflected shock.
This process may explain the X-ray morphology of this and even other
thermal composite supernova remnants.
\end{enumerate}

This work was supported by \Chandra\ grant GO3-4074X. 
Y.~C.\ acknowledges support from National Natural Science Foundation
of China grants 10725312, 10673003, and 10221001.
F.~D.~S.\ thanks Scott Wolk for useful advice on star identification
and Y.C.\ also thanks Jasmina Lazendic and Pat Slane for previous
collaboration on issues regarding reflected shock waves.
We also thank an anonymous referee for helpful suggestions.

\end{document}